\documentclass[traditabstract,times]{aa} % for the abstract without structuration 
\newcommand{\ergps}{erg\thinspace s$^{-1}$}
\newcommand{\phpspsqcm}{ph\thinspace cm$^{-2}$\thinspace s$^{-1}$}
\newcommand{\ergpspsqcm}{erg\thinspace s$^{-1}$\thinspace cm$^{-2}$}
\newcommand{\psqcm}{cm$^{-2}$}
\newcommand{\nH}{$N_{\rm H}$}

\usepackage{graphicx}
\begin{document}
\title{Testing a double AGN hypothesis for Mrk 273}

%   \subtitle{I. Overviewing the $\kappa$-mechanism}

   \author{K. Iwasawa\inst{1,2}\thanks{Email: kazushi.iwasawa@icc.ub.edu}
          \and
          Vivian U\inst{3,4,5}
          \and
          J.~M. Mazzarella\inst{6}
          \and
          A.~M. Medling\inst{7,8,9}
          \and
          D.~B. Sanders\inst{10}
          \and
          A.~S. Evans\inst{11,12}
%\fnmsep\thanks{Just to show the usage
%          of the elements in the author field}
}

\institute{Institut de Ci\`encies del Cosmos (ICCUB), Universitat de Barcelona (IEEC-UB), Mart\'i i Franqu\`es, 1, 08028 Barcelona, Spain
         \and
ICREA, Pg. Llu\'is Companys 23, 08010 Barcelona, Spain
\and
Department of Physics and Astronomy, 900 University Avenue, University of California, Riverside CA 92521, USA
\and
Department of Physics and Astronomy, 4129 Frederick Reines Hall, University of California, Irvine, CA 92697
\and
University of California Chancellor's Postdoctoral Fellow
\and
IPAC, MS 100-22, California Institute of Technology, Pasadena, CA 91125, USA
\and
Cahill Center for Astronomy and Astrophysics, California Institute of Technology, MS 249-17, Pasadena, CA 91125, USA
\and
Research School of Astronomy \& Astrophysics, Australian National University, Canberra, ACT 2611, Australia
\and
Hubble Fellow
\and
Institute for Astronomy, University of Hawaii, 2680 Woodlawn Drive, Honolulu, HI 96822, USA
\and
Department of Astronomy, University of Virginia, Charlottesville, VA 22903-2325, USA
\and
National Radio Astronomy Observatory, 520 Edgemont Road, Charlottesville, VA 22903-2475, USA
          }

%   \date{;}

% \abstract{}{}{}{}{} 
% 5 {} token are mandatory
 
\abstract{The ULIRG Mrk 273 contains two infrared nuclei, N and SW,
  separated by 1 arcsec. A Chandra observation has identified the SW
  nucleus as an absorbed X-ray source with \nH $\sim 4\times 10^{23}$
  \psqcm\ but also hinted at the possible presence of a Compton thick
  AGN in the N nucleus, where a black hole of $\sim 10^9 M_{\odot}$ is
  inferred from the ionized gas kinematics. The intrinsic X-ray
  spectral slope recently measured by NuSTAR is unusually hard
  ($\Gamma\sim 1.3$) for a Seyfert nucleus, for which we seek an
    alternative explanation. We hypothesise a
  strongly absorbed X-ray source in N, of which X-ray emission rises
  steeply above 10 keV, in addition to the known X-ray source in SW,
  and test it against the NuSTAR data, assuming the standard
    spectral slope ($\Gamma = 1.9$). This double X-ray source model
  gives a good explanation of the hard continuum spectrum, the deep Fe
  K absorption edge, and the strong Fe K line observed in this ULIRG,
  without invoking the unusual spectral slope required for a single
  source interpretation. The putative X-ray source in N is found to be
  absorbed by \nH $=1.4^{+0.7}_{-0.4}\times 10^{24}$ \psqcm. The
  estimated 2-10 keV luminosity of the N source is $1.3\times 10^{43}$
  \ergps, about a factor of 2 larger than that of SW during the NuSTAR
  observation. Uncorrelated variability above and below 10 keV between
  the Suzaku and NuSTAR observations appears to support the double
  source interpretation. Variability in spectral hardness and Fe K
  line flux between the previous X-ray observations is also consistent
  with this picture.}

\keywords{X-rays: galaxies - Galaxies: active - Galaxies: individual: Mrk 273
                             }
\titlerunning{Double X-ray source in Mrk 273}
\authorrunning{K. Iwasawa et al.}
   \maketitle
%
%________________________________________________________________

\section{Introduction}

Major mergers of gas-rich galaxies appear to play an important role in
the creation of ultra-luminous infrared galaxies (ULIRGs, Sanders \&
Mirabel 1996). Dense molecular gas channeled into the nuclear region
in the process of a merger drives intense star formation (Barnes \&
Hernquist 1992). Numerical simulations predict that the nuclear black
holes of merging galaxies will accrete gas simultaneously
(e.g. Hopkins et al 2005). Those active black holes (active galactic
nuclei or AGN) are naturally obscured heavily during the ULIRG
phase. As a result, they are sometimes difficult to identify due to
the nuclear obscuration as well as the surrounding star formation
activity that masks their AGN signatures.

Mrk 273 is a nearby ULIRG ($L_{\rm ir} = 10^{12.21} L_{\odot}$) at $z
= 0.038$, exhibiting a long tidal tail (Fig. 1), a clear signature of
a recent galaxy merger. There are three radio components, N, SW and SE
(Condon et al 1991) in the nuclear region (Fig. 1), two of which (N
and SW) coincide with near-infrared nuclei with a separation of 1
arcsec imaged by HST/NICMOS (Scoville et al 2000).

SW is a point-like, red near-infrared source and identified as an
absorbed hard X-ray source (Iwasawa et al 2011) with \nH $\simeq
4\times 10^{23}$ \psqcm\ (Iwasawa 1999; Xia et al 2002; Ptak et al
2003, Balestra et al 2005; Iwasawa et al 2011). Together with the
optical emission-line characteristics (Colina et al 1999), SW is
likely to contain an obscured Seyfert nucleus. However, much of the
infrared luminosity of Mrk 273 must originate from the other nucleus N
(see Iwasawa et al 2011 and references therein), where a concentration
of large amount of molecular gas is found (Downes \& Solomon 1998) and
intense star formation is taking place (Carilli \& Taylor 2000; Bondi
et al 2005). While the starburst appears to dominate the power output,
near-infrared integral-field spectroscopy has revealed a strongly
rotating disc in N, from which a black hole of $10^9 M_{\odot}$ is
inferred (U et al 2013, see also Kl\"ockner \& Baan 2004,
Rodor\'iguez-Zaul\'in et al 2014)). Given the environment of the N
nucleus, the black hole is expected to be buried deeply in the
dust/gas shrouds.

% Fig. 1 HST image N, SW, SE
\begin{figure}
\centerline{\includegraphics[width=0.48\textwidth,angle=0]{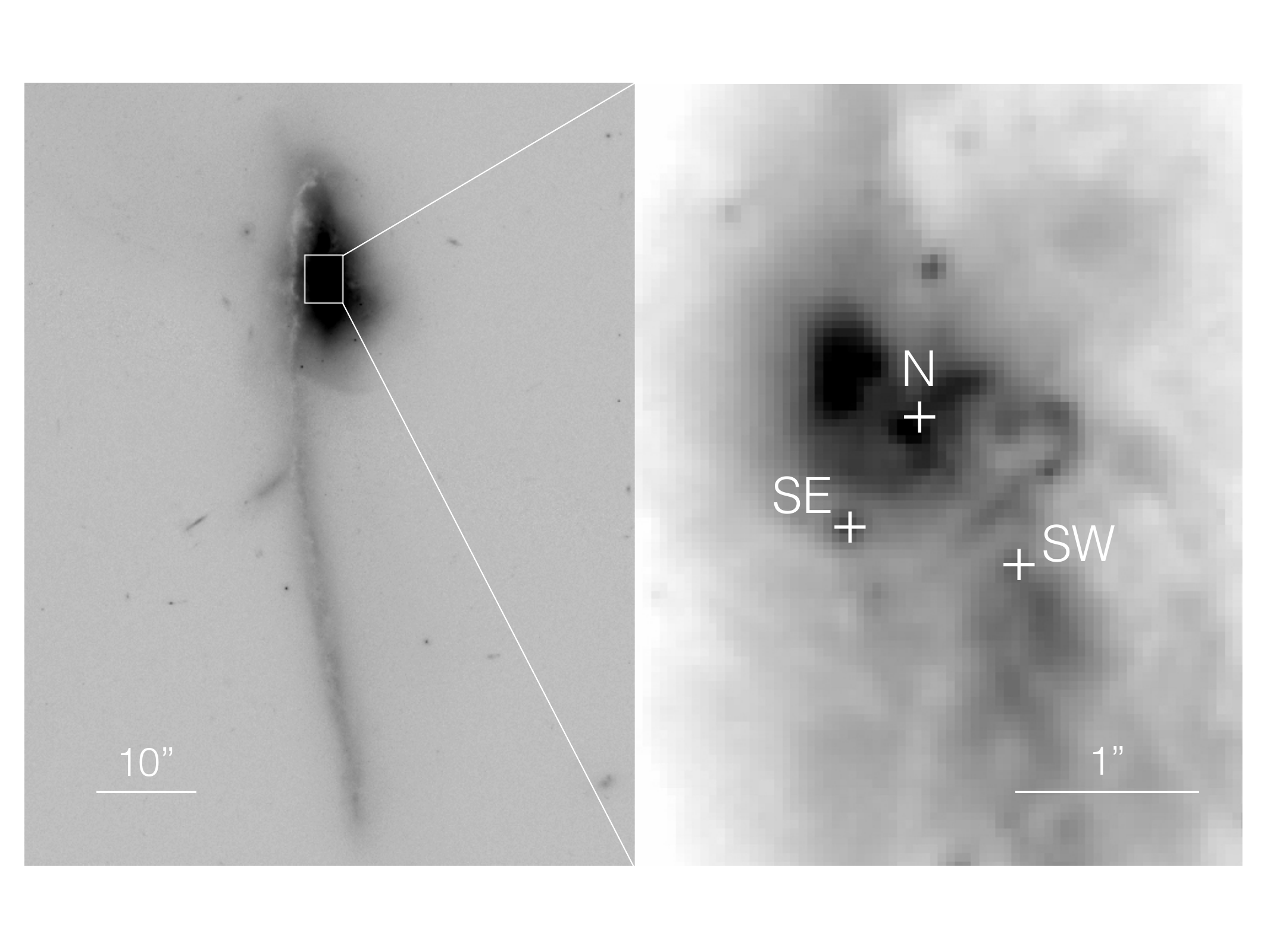}}
\caption{(Left) HST/ACS {\sl I} (F814W) band image of Mrk 273. The
  image orientation is north up, east to the left. The box indicates
  the zoomed region in the right panel. (Right) The nuclear region of
  Mrk 273. Positions of the three radio components, N, SW and SE
  (Condon et al 1991) are marked by plus symbols. Much of the
  far-infrared emission arises from the N nucleus while Chandra
  imaging found the SW nucleus to be a primary hard X-ray source
  (see Iwasawa et al 2011 for X-ray source identification).}
\end{figure}

Although SW dominates the Chandra hard X-ray (3-7 keV) image, the 6-7
keV narrow-band image containing the Fe K line (6.4 keV) shows a
possible extension towards N. This is in contrast to the point-like,
continuum-only (4-6 keV) image centred on SW, and points to N being
the origin of the enhanced Fe K line (Iwasawa et al 2011). The Chandra
data quality was insufficient to be conclusive, but if this is
true, it suggests presence of an X-ray source that exhibits a strong
Fe K line and faint underlying continuum in the N nucleus. These
characteristics fit an X-ray spectrum of a Compton thick AGN (with \nH
$\geq 10^{24}$ \psqcm), of which hard X-ray emission may rise above 10
keV.

The recently published NuSTAR observation of Mrk 273 (Teng et al 2015)
detected hard X-ray emission up to 30 keV. They estimated the spectral
slope of the absorbed continuum of $\Gamma\simeq 1.4$. This slope is
corrected for absorption and the effects of Compton scatterings in the
obscuring medium. A simple absorbed power-law fit (as usually used in
a spectral analysis of an X-ray survey) gives a harder slope of
$\Gamma\simeq 1.3$ (Sect. 2). It is rare to find such a hard {\it
  intrinsic} X-ray slope among AGN.  The photon index falls just below
the entire photon index distribution measured for the Swift/BAT hard
X-ray selected AGN and lies outside the $3 \sigma $ region of the
intrinsic scatter around the mean photon index ($\Gamma = 1.94$ for
Type 1 AGN and $\Gamma = 1.84$ for Type 2 AGN, Ueda et al
2014). Theoretically, extreme conditions are required, to produce a
power-law X-ray slope harder than $\Gamma = 1.5$ in the popular
disc-corona model (Haardt \& Maraschi 1993). Here, a single absorbed
source, the SW nucleus, is assumed to account for the entire NuSTAR
spectrum, as the double nucleus cannot be resolved spatially by X-ray
telescopes except Chandra. Suppose, in addition to the SW source, a
Compton thick AGN is present in N, which elevates the hard X-ray
emission above 10 keV, offering a natural explanation for the hard
NuSTAR spectrum.  We explore this possibility of a double X-ray source
assuming the standard X-ray spectral parameters usually found for AGN,
namely, a photon index $\Gamma = 1.9$ and solar metallicity.

The cosmology adopted here is $H_0=70$ km s$^{-1}$ Mpc$^{-1}$,
$\Omega_{\Lambda}=0.72$, $\Omega_{\rm M}=0.28$.

\section{NuSTAR spectrum}

% Fig. 2 NuSTAR spectrum fits APL and Etorus
\begin{figure}
\centerline{\includegraphics[width=0.5\textwidth,angle=0]{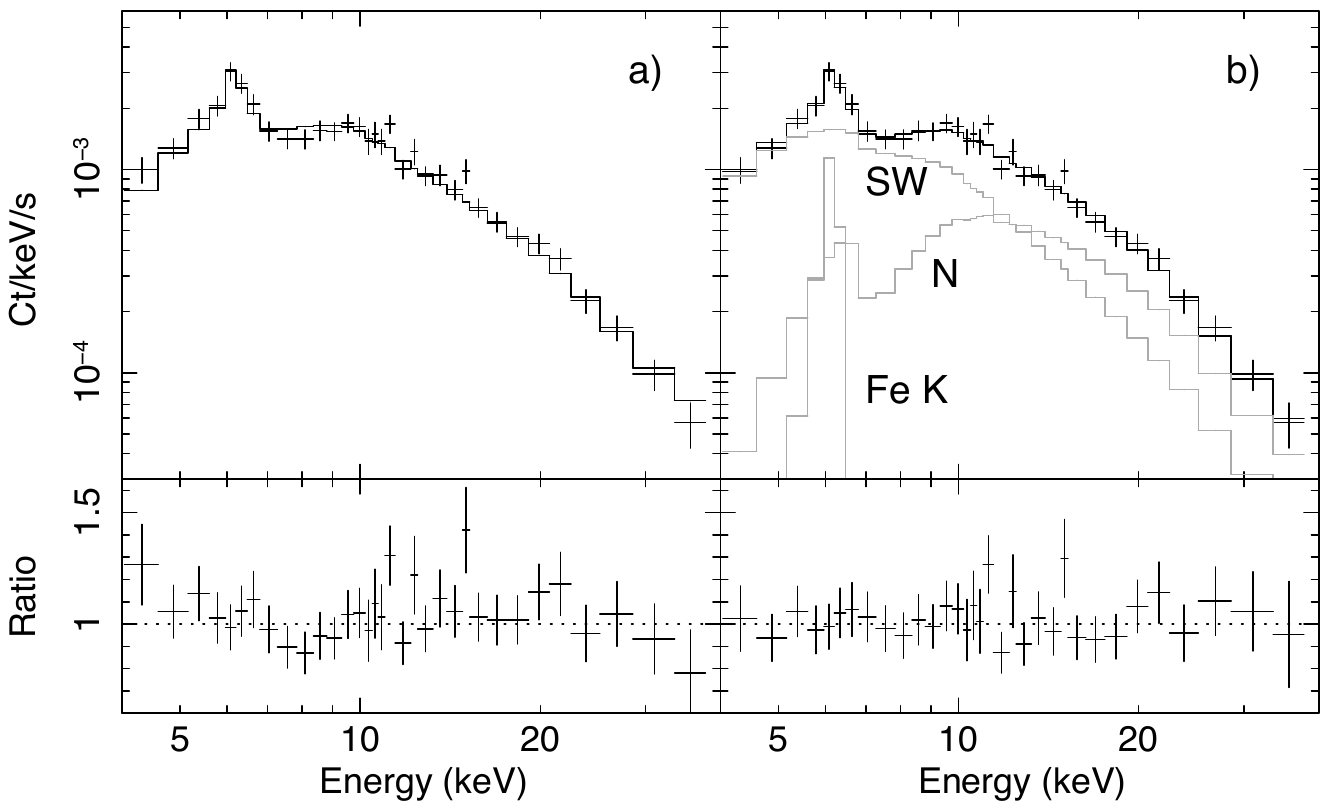}}
\caption{a) The NuSTAR count rate spectrum of Mrk 273 plotted with the
  best-fit absorbed power-law model (Model (i) of Table 1 in
  histogram). The model includes a Gaussian line for the Fe K line at
  6.4 keV. The data to model ratio is plotted in the bottom panel. The
  best-fit slope is very hard ($\Gamma = 1.26$). b) The same data
  plotted with the two source model, Model (iii). The best-fit models
  for the two absorbed continuum sources (labeled as SW and N) and the
  total Fe K line (labeled as Fe K) are drawn in light grey histograms. }
\end{figure}

% Table 1. NuSTAR fits
\begin{table}
\begin{center}
\caption{Spectral fits to the NuSTAR data}
\begin{tabular}{cccc}
\multicolumn{4}{c}{(i) Absorbed power-law}\\
$\Gamma $ & \nH & & $\chi^2$/dof\\
$1.26\pm 0.10$ & $3.0\pm 0.6$ & --- & 56.70/55\\[5pt]
\multicolumn{4}{c}{(ii) With enhanced Fe metallicity}\\
$\Gamma $ & \nH & $N_{\rm H}$(Fe) & $\chi^2$/dof\\
$1.41\pm 0.11$ & $3.2\pm 0.5$ & $4.5\pm 0.9$ & 48.80/54\\[5pt]
\multicolumn{4}{c}{(iii) SW \& N components}\\
$\Gamma $ & $N_{\rm H}^{\rm SW}$ & $N_{\rm H}^{\rm N}$ & $\chi^2$/dof\\
1.9 & $2.6^{+0.9}_{-1.1}$ & $14^{+7}_{-4}$ & 46.20/54\\
\end{tabular}
\begin{list}{}{}
\item[] Note --- Model (i) is a single power-law modified by cold
  absorption. $\Gamma $ is the photon index and \nH\ is the hydrogen
  equivalent column density in units of $10^{23}$ \psqcm, measured in
  the galaxy rest-frame. Model (ii) is the same as Model (i), except
  that \nH\ for Fe is fitted independently from the other
  elements. Model (iii) assumes both SW and N sources have the
  standard power-law slope $\Gamma = 1.9$ and are modified by
  different absorbing columns. Unlike in Model (i) and (ii), the
  effects of Compton scattering within the respective absorbing tori,
  computed by {\tt e-torus} (Ikeda et al 2009), are included (see text
  for details). A narrow Gaussian for the Fe K line at 6.4 keV is
  included in all the fits. The fits with Model (i) and (iii)
  correspond to Fig. 2a and Fig. 2b, respectively.
\end{list}
\end{center}
\end{table}

% Table 2 SW and N flux and luminosities
\begin{table}
\begin{center}
\caption{Decomposed fluxes and luminosities.}
\begin{tabular}{ccccc}
Nucleus& $F_{4-8}$ & $F_{10-30}$ & $L_{2-10}$ & $L_{10-30}$\\[5pt]
SW & 3.6 & 12 & 6.2 & 4.8 \\
N &  5.6 & 18 & 13 & 10 \\ 
\end{tabular}
\begin{list}{}{}
\item[] Note --- The estimated fluxes in the 4-8 keV and 10-30 keV
  bands in unit of $10^{-13}$ \ergpspsqcm, and the intrinsic
  luminosities in unit of $10^{42}$ \ergps, corrected for absorption
  and reflection for the SW and N nuclei, based on the two component
  decomposition (Model (iii) in Table 1 and Fig. 2b). The Fe K
  emission line flux is excluded from the measured fluxes.
\end{list}
\end{center}
\end{table}

NuSTAR observed Mrk 273 during guaranteed time on 2013 November 14
(ObsID 6000202802). We used cleaned event files processed by {\tt
  nupipeline} using the latest calibration files. The net exposure
time is 70 ks for both FPMA and FPMB modules. After checking that the
spectra from the two modules agree within errors, we used the averaged
spectrum for the spectral analysis below. The data above 4 keV are
used to avoid any extranuclear emission in the soft band. We binned
the spectrum so that each bin has more than 32 source counts and used
the $\chi^2$ minimisation method to search for best fits. The errors
quoted in each parameter are of $1 \sigma $.

First, the 4-40 keV spectrum is fitted by a power-law modified by a
cold absorber in the line of sight. Various effects of Compton
scattering in the absorbing medium, e.g., reflected continuum emission
when a torus-type geometry is assumed for an absorber, are ignored
here. This is to give an idea of the spectral shape and to make a
direct comparison of the result with the general properties of AGN
possible, as X-ray survey data are usually analysed in this way. A
narrow Gaussian is added to describe the 6.4 keV Fe K line. Fig. 2a
shows the NuSTAR data and the deviation from the best-fitting absorbed
powerlaw model, which finds $\Gamma= 1.26\pm 0.10$ and \nH $= (3.0\pm
0.6)\times 10^{23}$ \psqcm (as measured in the galaxy rest frame). The
spectral slope is remarkably hard, which primarily led us to seek an
alternative explanation with the double nucleus, as argued in
Sect. 1. The Fe K line is found at rest energy of $6.39\pm 0.06$ keV
with intensity of $(4.6\pm 1.0)\times 10^{-6}$ \phpspsqcm. The
equivalent width (EW) of the line to the absorbed continuum is
$0.36\pm 0.08$ keV. We note that this model systematically
overestimates the data in the 7-9 keV range, where the Fe K absorption
edge (the threshold energy at 7.1 keV for neutral Fe) is present. The
above fit of the absorbing column is primarily driven by the continuum
rollover at the low energy side. The mismatch of the Fe K absorption
band indicates that the Fe absorption edge is deeper than predicted by
the best-fit \nH. This can be explained by increased Fe metallicity
(relative to the solar abundance of Anders \& Grevesse 1989). When Fe
K metallicity is fitted (while other elements are tied together and
independent from Fe), $N_{\rm H}$(Fe) is found to be $(4.5\pm
0.9)\times 10^{23}$ \psqcm, while \nH $=(3.2\pm 0.5)\times 10^{23}$
\psqcm\ for the other elements. Although statistical evidence for
excess Fe metallicity is not strong, it is suggestive for the proposed
two component model as discussed below.

In Sect. 1, we mentioned the possible presence of a Compton thick AGN in
the N nucleus in addition to the SW nucleus, suggesting that the hard NuSTAR
continuum as measured above may result from elevated continuum
emission above 10 keV due to a strongly absorbed source in N. Here, a
surmised N source would have \nH $\sim 10^{24}$ \psqcm. The above
inspection of the NuSTAR spectrum gives two more properties to favour
this hypothesis: the deep Fe K absorption edge and the large EW of the Fe K
line (although they are not as strong as the spectral slope). An
additional strongly absorbed source would enhance the total Fe edge
depth without invoking the excess Fe metallicity. The observed Fe line
EW of 0.36 keV is larger than expected for an absorbing torus with \nH
$\sim 3\times 10^{23}$ \psqcm (EW $\simeq 0.2$ keV, Ikeda et al 2009;
Ghisellini, Haardt \& Matt 1994; Awaki et al 1991). In fact, the
spectral fit using the {\tt MYTORUS} model presented in Teng et al
(2015) appears to underestimate the Fe K line flux (their
Fig. 5). Extra line emission from N, as hinted by the Chandra
observation (see Sect. 1), could account for the excess line emission.

We here proceed to model the NuSTAR spectrum with two absorbed nuclei,
SW and N. The N nucleus is hypothesised to have an X-ray source
absorbed by a larger \nH\ than that for SW. We assume that each
nucleus has an X-ray source with the standard power-law slope of
$\Gamma = 1.9$ (e.g. Nandra \& Pounds 1994; Ueda et al 2014). Effects
of reflection within the respective absorbers computed by Monte Carlo
simulations assuming a torus geometry (e-torus: Ikeda et al 2007;
details of the usage are described in Awaki et al 2009) are
included. Fe K line emission, which is not included in {\tt e-torus},
is modelled by a Gaussian, which accounts for a sum of the line
emission from both nuclei. Since the inclination and opening angles of
the tori of the respective nuclei cannot be constrained independently,
we made the following assumptions of the torus geometry: both tori are
inclined at $55^{\circ}$ (which is the inclination angle measured for
the gas disc at N by U et al 2013 but not known for SW); the opening
angles are $50^{\circ}$ for SW and $10^{\circ}$ for N, as a relatively
large opening of the torus is expected for SW exhibiting Seyfert 2
characteristics (Colina et al 1999; U et al 2013) while a small
opening for the dense-gas filled N nucleus. Solar metallicity is
assumed for both absorbers.

This double source model describes the data well (Fig. 2b), with the quality of
fit comparable to or even better than those single source models (Table 1). The
model agrees with the Fe K absorption band well. The assumed steep
spectral slope fits better the data at the high energy end (25-40
keV), which results in a slightly smaller $\chi^2$ than in Model
(ii). The absorbing columns are found to be \nH $=
2.6^{+0.9}_{-1.1}\times 10^{23}$ \psqcm\ for SW and \nH $=
1.4^{+0.7}_{-0.4}\times 10^{24}$ \psqcm\ for N. The fit suggests that
the hypothetical N nucleus is absorbed by a Thomson depth of unity, in
agreement with a suspected Compton thick AGN. According to the fit, 87
per cent of the observed 4-8 keV flux comes from the SW component
while the N component, which is faint at lower energies, rises sharply
towards 10 keV and exceeds SW at $\sim 12$ keV upwards (see
Fig. 2b). The observed 4-8 keV and 10-30 keV fluxes and the intrinsic
continuum source luminosities in the 2-10 keV bands for the SW and N
components, estimated from the fit, are given in Table
2. Intrinsically, the hidden N nucleus appears to be a factor of 2
more luminous than the SW nucleus.

The measured Fe K line flux found in the above fit is $(4.1\pm
1.0)\times 10^{-6}$ \phpspsqcm, with contributions from both
nuclei. According to Ikeda et al 2009, EW $= 0.18$ keV is expected for
an illuminated torus with \nH $=2.6\times 10^{23}$ \psqcm, as obtained
for SW. The corresponding line intensity is $2.2\times 10^{-6}$
\phpspsqcm. The rest of the line flux ($(1.9\pm 1.0)\times 10^{-6}$
\phpspsqcm) can then be attributed to N. The corresponding EW with
respect to the N continuum is $EW\simeq 0.7$ keV, which is expected
for \nH $\simeq 1\times 10^{24}$ \psqcm, in agreement with the
estimated \nH\ for N.  The same spectral fit using
absorption/reflection spectra with Fe line emission, computed by other
codes, e.g., {\tt MYTORUS} (Murphy \& Yaqoob 2009) instead of {\tt
  e-torus} gives a consistent result.

The spectrum taken from the region of the N nucleus in the earlier
Chandra observation shows marginal evidence of a strong 6.4 keV line
(Iwasawa et al 2011) with intensity of $1.4\pm 0.7$ \phpspsqcm, which
is comparable to the above estimate of the Fe line in N based on the
NuSTAR. In addition to emission from a putative AGN, the strong
starburst taking place in N could generate hot gas emitting
high-ionization Fe (mainly Fe {\sc xxv}) emission, which may
contributes to the 6-7 keV image extension towards N seen by
Chandra. There are strong cases for such high-ionization Fe emission
in U/LIRGs, Arp220 (Iwasawa et al 2005, 2009; Teng et al 2009, 2015)
and NGC 3690E (Ballo et al 2004; Ptak et al 2015). Assuming the
high-ionization Fe emission is proportional to star formation rate, we
estimated its line flux, based on the measurements for Arp 220 and NGC
3690E, taking into account the source distances. It is found to be
$(4-8)\times 10^{-7}$ \phpspsqcm, approximately 10-20\% of the total
line flux detected with NuSTAR.

\section{Spectral variability}

% Fig. 3 Four panel low-reso spectra
\begin{figure*}
\centerline{\includegraphics[width=0.8\textwidth,angle=0]{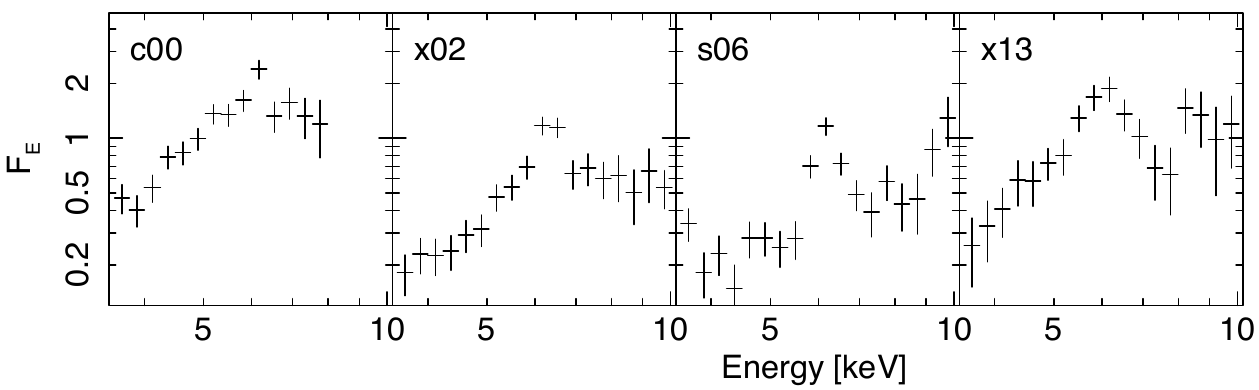}}
\caption{The X-ray spectra of Mrk 273 observed at the four epochs
  (Table 3). The data are plotted in flux density unit of $10^{-13}$
  erg~s$^{-1}$~cm$^{-2}$~keV$^{-1}$. }
\end{figure*}

% Fig. 4 Suzaku - Nustar spectra in one panel
\begin{figure}
\centerline{\includegraphics[width=0.35\textwidth,angle=0]{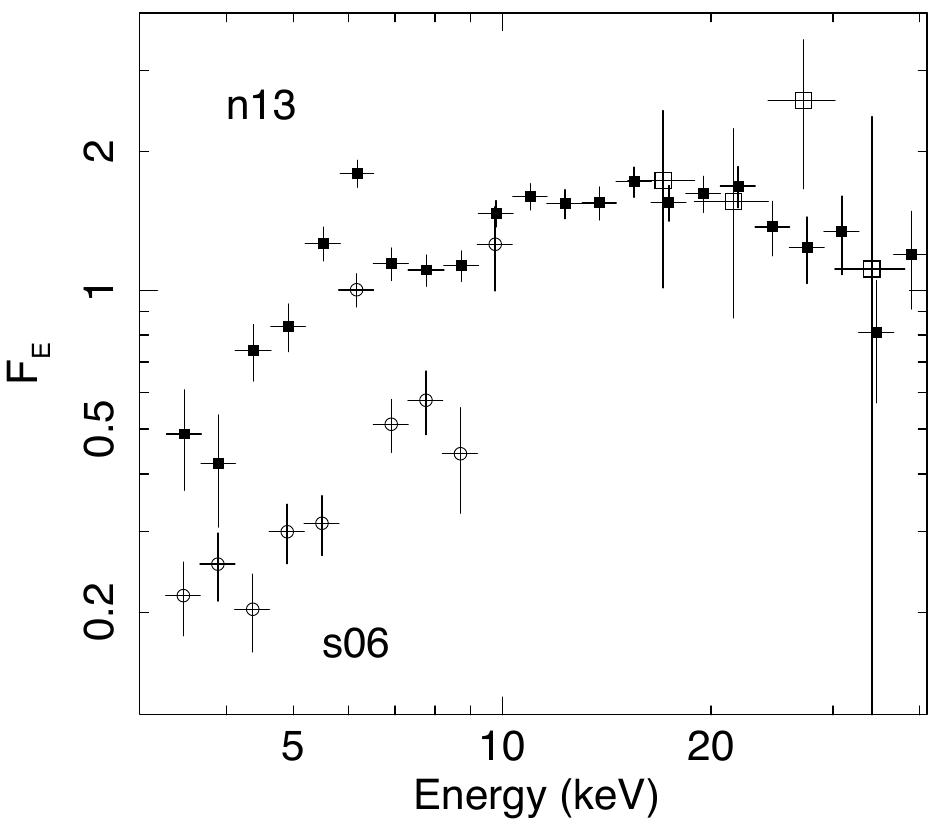}}
\caption{Same as Fig. 3 but the Suzaku XIS (open circles) and PIN
  (open squares) data from s06 are plotted with the NuSTAR broad-band data (filled
  squares). }
\end{figure}

% Fig. 3 F4-8 vs Fe line correlation
\begin{figure}
  \centerline{\includegraphics[width=0.35\textwidth,angle=0]{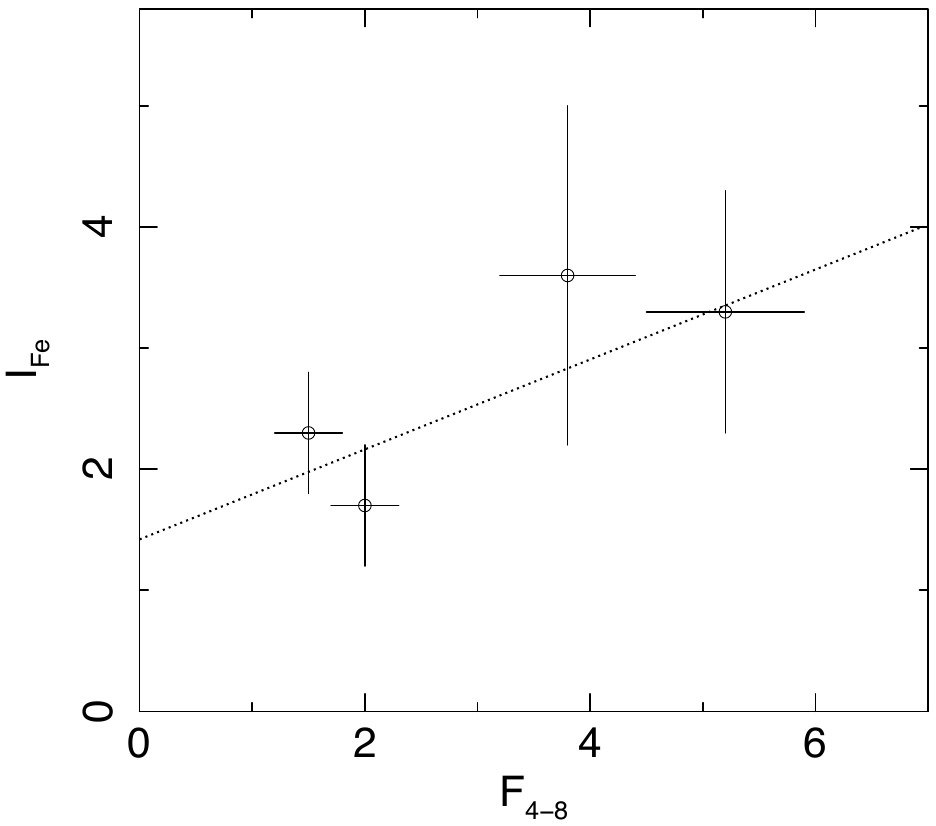}}
  \caption{Correlation diagram of the Fe K line and 4-8 keV continuum
    fluxes between the four observations. The 4-8 keV continuum flux
    and the Fe line intensity are in units of $10^{-13}$
    \ergpspsqcm\ and $10^{-6}$ \phpspsqcm, respectively. }
\end{figure}

% Observation log
\begin{table*}
\begin{center}
\caption{X-ray observations of Mrk 273}
\begin{tabular}{clcrcccc}
Observation & Observatory & Date & ObsID & Detector & Exposure & Counts & Band \\
& & yyyy-mm-dd & & & ks & ct s$^{-1}$\\[5pt]
c00 & Chandra & 2000-04-19 & 809 & ACIS-S & 45 & 533 & 4-8\\
x02 & XMM-Newton & 2002-05-07 & 0101640401 & EPIC & 18/22 & 573 & 4-10\\
s06 & Suzaku & 2006-07-07 & 701050010 & XIS/PIN & 80/77 & 626/714 & 4-10/15-40\\
x13 & XMM-Newton & 2013-11-04 & 0722610201 & EPIC & 3.5/20 & 478 & 4-10\\
n13 & NuSTAR & 2013-11-04 & 6000202802 & FPMA/B & 70 & 2930 & 4-40\\
\end{tabular}
\begin{list}{}{}
\item[] Note --- XMM-Newton carries three EPIC cameras (pn, MOS1 and
  MOS2). The exposure times for the pn and two MOS cameras are shown
  separately: (pn)/(average of MOS1 and MOS2), but the source counts
  are from all the three cameras. We use Suzaku data from the four XIS
  cameras and the HXD-PIN. The exposure times, source counts and
  energy bands in which the source counts were measured are given
  separately. All the XIS source counts are summed together.
\end{list}
\end{center}
\end{table*}

% Table 4 Spectral changes
\begin{table}
\begin{center}
\caption{Variability between four obserations.}
\begin{tabular}{cccc}
Obs & \nH & $I_{\rm Fe}$ & $f_{4-8}$ \\
(1) & (2) & (3) & (4) \\[5pt]
c00 & $2.9\pm 0.5$ & $3.3\pm 1.0$ & $5.2\pm 0.7$ \\
x02 & $3.8\pm 0.5$ & $1.7\pm 0.5$ & $2.0\pm 0.3$ \\
s06 & $3.5\pm 0.7$ & $2.3\pm 0.5$ & $1.5\pm 0.3$ \\
x13 & $3.1\pm 0.6$ & $3.6\pm 1.4$ & $3.8\pm 0.6$ \\
%x+n13 & $3.0\pm 0.3$ & $3.6\pm 1.4$ & $4.1\pm 0.3$ \\
\end{tabular}
\begin{list}{}{}
\item[] Note --- (1) The observations indicated in Table 1; (2)
  Absorbing column density \nH\ measured in the galaxy rest frame in
  units of $10^{23}$ \psqcm; (3) Iron line intensity $I_{\rm Fe}$ in
  units of $10^{-6}$ \phpspsqcm, as observed, without correction for
  absorption; (4) Source flux in the 4-8 keV band in units of
  $10^{-13}$ \ergpspsqcm. The flux is measured for the continuum (the
  Fe K line flux has been subtracted).
\end{list}
\end{center}
\end{table}

There are previous X-ray observations of Mrk 273 mainly at energies lower
than 10 keV, and the source flux is variable. According to the spectral
decomposition, while the observed continuum flux below 10 keV is dominated by
SW, the two nuclei may have comparable contributions to the Fe K
line. Spectral variability, including the Fe line, accompanied by
total X-ray brightness changes may give a clue to the reality of the
two component modelling. We use four other X-ray observations of Mrk
273 performed with the Chandra X-ray Observatory (Chandra), Suzaku, and
XMM-Newton between 2000 and 2013 (Table 3, they are labeled as c00,
x02, s06 and x13). There are two XMM-Newton observations, one of which
(x13) was carried out simultaneously with the NuSTAR observation
(n13). The 4-10 keV data of the EPIC pn and two MOS cameras of
XMM-Newton and the Suzaku XIS and the 4-8 keV data of Chandra were
used. Their spectra are shown in Fig. 3. The 4-8 keV X-ray source flux
observed in c00 dropped significantly during x02 and s06, and then
recovered during x13 and n13.

Among these previous observations, marginal detection in the 15-40 keV
band with the Suzaku HXD-PIN during s06 (Teng et al 2009) may have
potential significance to our hypothesis. Our own analysis (with the
2009-Aug version of {\tt hxdpinxbpi} and the ``tuned'' ver. 2.0
background) obtained 15-40 keV flux of $(4.3\pm 1.1)\times 10^{-12}$
\ergpspsqcm (the error is statistical uncertainly only) and the
spectral data are plotted in Fig. 4, along with the 3-10 keV XIS data
taken in the same observation and the NuSTAR data. The source excess
($\sim 3$\%) is slightly larger than the systematic uncertainty of the
background ($\sim 1.3$\% in 15-40 keV, Mizuno et al 2008) and the
detection remains marginal. Supposing this PIN measurement is real,
Fig. 4 shows that the observed 15-40 keV flux is comparable between
s06 and n13 ($f_{15-40}=(3.3\pm 0.3)\times 10^{-12}$ \ergpspsqcm),
whilst the Suzaku XIS flux below 10 keV is more than a factor of 2
fainter than that of NuSTAR (see Table 4). This uncorrelated
variability above and below 10 keV suggests two distinct sources are
present in the respective bands (subject to the reliability of the PIN
detection), in agreement with the NuSTAR spectral decomposition
(Fig. 2b). In the double source picture, the SW source is variable
while the N source maintains similar brightness. The same spectral
model fitted to the NuSTAR data (Model (iii) of Table 1) gives a good
description of the Suzaku spectrum by changing only the continuum
normalization for SW and the Fe line flux.

Now we look at spectral variability below 10 keV between the four
observations by a simple modelling. All the spectra are modelled by an
absorbed power-law plus a narrow Gaussian for the Fe K line. We assume
a power-law slope of $\Gamma = 1.26$ obtained from the NuSTAR spectrum
(Sect. 2). The column density of cold absorber \nH, power-law
normalization and Fe line strength are left variable between the
observations. If SW is solely responsible for the observed X-ray
emission, the measured \nH\ is the absorbing column towards the SW
nucleus, while in the context of the double source model, it is a
measure of spectral hardness in the 4-10 keV band of the summed
spectrum of both nuclei. We note that the spectral shown in Fig. 3
have been corrected for the detector response curve (Iwasawa et al
2012) and detector-averaged when data from multiple detectors are
available (XMM-Newton EPIC and Suzaku XIS) for convenience of a visual
comparison. Spectral fitting was performed for uncorrected data from
the individual detectors jointly, using the detector responses and the
background, to obtain the spectral parameters of interest. The
absorbing column density, Fe line intensity and the 4-8 keV flux
estimated from the spectral fits are shown in Table 4.

Among the four observations, the continuum source is bright in c00 and
x13 and faint in x02 and s06. The variability results on \nH\ and Fe
line flux between the bright and faint states appear to be more
consistent with the ``SW + N'' hypothesis than that of SW alone, as
explained below.

The Fe line flux dropped when the 4-8 keV continuum flux was low (x02
and s06) and rose back when the continuum brightened (x13),
suggesting that the line responds to the continuum
illumination. However, the correlation plot (Fig. 5) shows a positive
offset in line flux at zero continuum intensity. As the SW nucleus
dominates the observed 4-8 keV continuum, the continuum variability
should be driven by SW. The responding Fe line is therefore attributed
to that from SW and the offset line flux ($\simeq 1.4\times
10^{-6}$ \phpspsqcm ) can be considered to originate from N. This line
flux decomposition roughly agrees with that from the EW argument for
the NuSTAR data (Sect. 2). We note that this interpretation relies on the
condition that the Fe line emitter in SW lies sufficiently close to
the central source so that it follows the continuum variability within
the intervals between the observations (a few years, corresponding to
a $\sim $pc in physical scale).

On the other hand, the inferred \nH\ is larger when the 4-8 keV flux
is low (Table 4). In the dual AGN scenario, this would be a natural
consequence. When the 4-8 keV flux (which is largely due to SW)
lowered, a relative contribution of the more strongly absorbed N
became larger (as N appears to remain in similar flux level, as
discussed above, see Fig. 4), resulting in a harder spectrum which
manifests itself as a larger absorbing column as measured. Here we
assume that \nH\ of the two nuclei remain the same which may not
necessarily hold. The variability of \nH\ in absorbed AGN is not
uncommon (e.g. Risaliti, Elvis \& Nicastro 2002). However, since the
absorbing clouds in the line of sight do not know what the continuum
source does, \nH\ variability should be independent of the continuum
flux and any correlated behaviour with continuum brightness as seen
here would not exist. A larger number of measurements are needed to
test the randomness of \nH\ variability.

In summary, the disconnected flux variability below and above 10 keV
seen between s06 and n13 appears to support two distinct
sources, as inferred by the NuSTAR spectral decomposition, indicating
that the SW source is significantly variable while the N source is
more stable between the observations. The Fe K line flux and the
spectral hardness, as a function of source brightness, also behave as
expected from the double source hypothesis, under reasonable
conditions.

\section{Discussion}

If the SW nucleus alone accounts for the X-ray emission of Mrk 273
observed with NuSTAR, some extreme conditions to produce the unusually
hard continuum slope and the excess Fe metallicity are required. For
the former, for instance, a compact corona (thus injection of soft
photons from the disc is minimized) with a large optical depth, e.g.,
$\tau\sim 10$, may be a requirement. Blazar-like jet emission, which
could produce a hard X-ray spectrum, is unlikely because SW is a weak
radio source (Condon et al 1991). While SW as the sole X-ray source
cannot be ruled out, the introduction of an additional, strongly
absorbed X-ray source, hypothesised in N, offers a natural explanation
for the NuSTAR spectrum with X-ray characteristics that are normal for
AGN (Sect. 2) as well as the various spectral variability (Sect. 3). U
et al (2013) also argued for a buried AGN in N, using the H$_2$
outflow energetics and the photoionization condition inferred from the
[Si {\sc vi}]/Br$\gamma $ ratio (see also Rodr\'iguez-Zaul\'in et al
2014), in support of the putative Compton thick AGN. It might be
possible to test the two source hypothesis/scenario with Fe K narrow
band imaging (Iwasawa et al 2011) of improved data quality (new 200 ks
Chandra data have been acquired recently, PI: S. Veilleux) but we
probably have to wait for imaging with a future high-resolution hard
X-ray telescope to be decisive.

The 2-10 keV absorption-corrected luminosity of the hypothetical N
nucleus is estimated to be $1.3\times 10^{43}$ \ergps, comparable to
Seyfert nuclei. The bolometric AGN luminosity of N is then $\sim
2\times 10^{44}$ \ergps\ when the bolometric correction of Marconi et
al (2004) is used. It is about two orders of magnitude below the total
infrared luminosity, indicating that the AGN does not dominate the
total energy production in Mrk 273. With a black hole mass of $1\times
10^9 M_{\odot}$, the Eddington ratio is about $10^{-3}$. NGC 6240 is a
well-known example of a double AGN system in LIRGs with two
Compton-thick nuclei (Komossa et al 2003). If proved true, the double
AGN in Mrk 273 would be similar to that found in other LIRGs, Mrk 266
(Mazzarella et al 2012) and IRAS 20210+1121 (Piconcelli et al 2010)
and with one nucleus being a Compton thick AGN.

\begin{acknowledgements}
This research made use of data obtained from NuSTAR, Chandra X-ray
Observatory, XMM-Newton and Suzaku, reduced and analysed by software
provided by HEASARC, the Chandra X-ray Center, and the XMM-Newton
Science Operation Centre. KI acknowledges support by the Spanish
MINECO under grant AYA2016-76012-C3-1-P and MDM-2014-0369 of ICCUB
(Unidad de Excelencia 'Mar\'ia de Maeztu'). VU acknowledges funding
support from the University of California Chancellor's Postdoctoral
Fellowship. Support for AMM is provided by NASA through Hubble
Fellowship grant \#HST-HF2-51377 awarded by the Space Telescope Science
Institute, which is operated by the Association of Universities for
Research in Astronomy, Inc., for NASA, under contract NAS5-26555.
\end{acknowledgements}

\end{document}